\documentclass[prb,twocolumn,showpacs,amsmath,amssymb,floatfix]{revtex4-1}
\usepackage{graphicx}
\usepackage{dcolumn}
\newcolumntype{d}[1]{D{.}{.}{#1}}
\usepackage{longtable}

\usepackage{epstopdf}

\def\beq{\begin{equation}}
\def\eeq{\end{equation}}

\newcommand{\qir}{Q_{\textrm{IR}}}
\newcommand{\qr}{Q_{\textrm{R}}}
\newcommand{\Tou}{T_{1u}}
\newcommand{\oml}{\Omega_{\textrm{l}}}
\newcommand{\omh}{\Omega_{\textrm{h}}}

\newcommand{\omhz}{\Omega_{\textrm{h}_z}}
\newcommand{\omlz}{\Omega_{\textrm{l}_z}}
\newcommand{\omlx}{\Omega_{\textrm{l}_x}}
\newcommand{\omly}{\Omega_{\textrm{l}_y}}
\newcommand{\qhx}{Q_{\textrm{h$_x$}}}
\newcommand{\qhz}{Q_{\textrm{h$_z$}}}
\newcommand{\qlx}{Q_{\textrm{l$_x$}}}
\newcommand{\qlz}{Q_{\textrm{l$_z$}}}
\newcommand{\qly}{Q_{\textrm{l$_y$}}}
\newcommand{\ampunit}{\AA $\sqrt{\textrm{amu}}$}
\newcommand{\zmstar}{Z^*_m}
\newcommand{\zmstarhx}{Z^*_{\textrm{h$_x$}}}
\newcommand{\zmstarhz}{Z^*_{\textrm{h$_z$}}}

\begin{document}

\title{Proposal for midinfrared light--induced ferroelectricity in oxide paraelectrics}

\author{Alaska Subedi} 

\affiliation{Centre de Physique Theorique, Ecole Polytechnique, CNRS,
  Universit\'e Paris-Saclay, F-91128 Palaiseau, France}
\affiliation{Coll\`ege de France, 11 place Marcelin Berthelot, 75005
  Paris, France}

\date{\today}

\begin{abstract}
  I show that a nonequilibrium paraelectric to ferroelectric
  transition can be induced using midinfrared pulses. This relies on a
  quartic $l\qlz^2\qhx^2$ coupling between the lowest ($\qlz$) and
  highest ($\qhx$) frequency infrared-active phonon modes of a
  paraelectric material. Density functional calculations show that the
  coupling constant $l$ is negative, which causes a softening of the
  $\qlz$ mode when the $\qhx$ mode is externally pumped. A
  rectification along the $\qlz$ coordinate that stabilizes the
  nonequilibrium ferroelectric state occurs only above a critical
  threshold for the electric field of the pump pulse, demonstrating
  that this is a nonperturbative phenomenon. A first principles
  calculation of the coupling between light and the $\qhx$ mode shows
  that ferroelectricity can be induced in the representative case of
  strained KTaO$_3$ by a midinfrared pulse with a peak electric field
  of 17 MV cm$^{-1}$ and duration of 2 ps. Furthermore, other
  odd-order nonlinear couplings make it possible to arbitrarily switch
  off the light-induced ferroelectric state, making this technique
  feasible for all-optic devices.
\end{abstract}

\pacs{77.80.Fm,78.20.Bh,63.20.Ry,78.47.J-}

\maketitle

\section{Introduction}

Living organisms have used light to observe the properties of
materials since the evolutionary development of complex eyes. However,
ultrafast light-control of materials properties has only become
feasible after the construction of high-powered lasers in the previous
century, and this field has flourished because light-induced processes
have the potential to lead to new devices and physical phenomena.
Many examples of light-induced phase transitions using near-visible
sources have been observed, including discontinuous volume changes in
polymer gels,\cite{suzu90} quasiionic-to-quasineutral transition in
organic molecular compounds,\cite{kosh90} low-spin to high-spin
transition in metal organic frameworks,\cite{decu84}
insulator-to-metal transition in perovskite manganites,\cite{miya97}
and, remarkably, a transition to a hidden metastable state in
$1T$-TaS$_2$.\cite{stoj14} All of these examples involve transition to
higher-temperature or metastable phases, and the quest to stabilize a
phase with less symmetry or more order using light remains elusive.

More recently, intense midinfrared pulses have been used to directly
control the dynamical degrees of freedom of the crystal lattice. Such
mode-selective vibrational excitations have been used to induce
insulator-to-metal transitions\cite{rini07,cavi12} and melting of
orbital\cite{tobe08} and magnetic\cite{fors11a,fors15} orders.
Midinfrared excitations have so far not caused transitions to
more-ordered phases, although light-induced superconductivity has been
claimed in several cuprate compounds and
K$_3$C$_{60}$.\cite{faus11,kais14,hu14,mitr16} However, these claims
rely on interpreting the two-dimensional response function
$\Sigma(\omega,\tau)$ measured in the pump-probe experiments as the
optical conductivity $\sigma(\omega)$ that is measured in time-domain
spectroscopy. It is unclear whether such an interpretation is
justified, especially at low frequencies, when the light-induced state
is short-lived, as is the case in these
experiments.\cite{kind99,oren15} In any case, a light-induced
transition to a lower-symmetry phase is not observed in any of these
experiments. Nonetheless, midinfrared excitation should be an
effective tool to drive materials to broken-symmetry phases because
selective and coherent excitation of the low-energy structural degrees
of freedom should cause minimal dissipation as heat.

The mechanism for mode-selective light-control of materials was
proposed by F\"orst \textit{et al.}\cite{fors11b,fors13} This involves
exciting an infrared-active phonon mode $\qir$ of a material using an
intense light pulse which then causes the lattice to displace along a
fully symmetric $A_g$ Raman mode coordinate $\qr$ due to a nonlinear
coupling $\qr\qir^2$ between the two modes. A quantitative microscopic
theory of this phenomenon was developed in Ref.~\onlinecite{sube14},
and calculations based on this theory in combination with a
time-resolved x-ray diffraction experiment was used to resolve the
midinfrared light--induced changes in the structure of
YBa$_2$Cu$_3$O$_{6.5}$.\cite{mank14} In addition to the historically
discussed cubic order $\qr Q_{\textrm{IR}_1} Q_{\textrm{IR}_2}$
coupling,\cite{wall71,mart74} Subedi \textit{et al.}\ have shown that
a sizeable quartic order $\qr^2\qir^2$ coupling can occur and studied
the dynamics due to such a coupling.\cite{sube14} They found that such
a quartic coupling exhibits various distinct regimes of dynamics,
including transient mode softening and dynamic stabilization in a
rectified state. In contrast to the case of the cubic coupling, the
displacement along the $\qr$ coordinate occurs only above a critical
pump amplitude threshold for the quartic $\qr^2\qir^2$ coupling. A
more recent work has reproduced several aspects of the dynamics of
this coupling.\cite{fech16} Unlike the cubic $\qr Q_{\textrm{IR}_1}
Q_{\textrm{IR}_2}$ coupling,\cite{jura16,q1q2q3} the $\qr^2\qir^2$
coupling can cause a rectification along a symmetry breaking
mode,\cite{sube14} but such a light-induced symmetry breaking of a
crystal structure has so far not been observed.

It has recently been predicted that ferroelectric polarization can be
switched using midinfrared pulses.\cite{sube15} In this paper, I
extend that work to the paraelectric phase and show that
ferroelectricity can also be induced in transition metal oxide
paraelectrics. This relies on a quartic $l\qlz^2\qhx^2$ coupling and
is a nonperturbative effect that occurs only above a critical pump
amplitude. Here, $\qhx$ is a high-frequency infrared-active phonon
mode that should be externally pumped and $\qlz$ is the lowest
frequency infrared-active mode that is transverse to the pumped
mode. I find that the sign of the coupling constant $l$ is negative in
several transition metal oxides, which causes the $\qlz$ mode to
soften as the $\qhx$ mode is externally pumped. But other quartic
order couplings $t_1 \qlx^3 \qhx$, $t_2 \qlx^2 \qhx^2$, and $t_3 \qlx
\qhx^3$ between $\qhx$ and the lowest frequency mode $\qlx$ that is
longitudinal to the pumped mode are larger in magnitude.  In the cubic
materials, the couplings between $\qlx$ and $\qhx$ modes are such that
the $\qlz$ mode may not develop a light-induced dynamical
instability. However, I find that the couplings in the longitudinal
direction can be effectively reduced by applying strain so that a
light-induced ferroelectric state is stabilized by rectification along
the $\qlz$ coordinate.

I illustrate this theory for the representative case of KTaO$_3$ to
show that light-induced ferroelectricity can be generated in the
strained version of this material when a pump pulse with an electric
field of $\sim$17 MV cm$^{-1}$ and pulse duration of 2 ps is
used. Interestingly, this value is noticeably smaller than what is
expected for the critical pump amplitude due to a $\qlz^2\qhx^2$
coupling.\cite{sube14} I find that this reduction is due to the
presence of substantial sixth order $\qlz^4\qhx^2$ and $\qlz^2\qhx^4$
couplings. Furthermore, I show that the light-induced rectification
can be arbitrarily suppressed by pumping the highest frequency
infrared-active mode $\qhx$ that is longitudinal to $\qlx$ with
another weak pulse. Such a control is necessary for applications in
devices. In addition to KTaO$_3$, I find similar nonlinear couplings
in SrTO$_3$, and LaAlO$_3$, and this technique could be generally
applied to many transition metal oxide paralectrics.

\section{Approach}

\subsection{Computational details}
The phonon frequencies and eigenvectors, nonlinear couplings between
different normal mode coordinates, and the coupling between light and
pumped infrared mode were all obtained from first principles using
density functional calculations as implemented in the {\sc vasp}
software package. I used the projector augmented wave pseudopotentials
provided with the package with the electronic configurations $3s^2
3p^6 4s^1$ (K), $5p^6 6s^2 5d^3$ (Ta), and $2s^2 2p^4$ (O, normal
cut-off). A plane-wave cut-off of 550 eV for basis-set expansion, an
$8 \times 8 \times 8$ $k$-point grid for Brillouin zone sampling, and
the PBEsol version of the generalized gradient approximation was
used.\cite{pbesol}

The calculations were done using the relaxed lattice parameters for
the cubic structure. For the strained structure, the $c$ lattice
parameter that minimized the total energy for the given strain was
used. A very small energy convergence criteria of $10^{-8}$ eV was
used in the calculations to ensure high numerical accuracy. After
relaxing the lattice parameters, I calculated the phonon frequencies
and eigenvectors using the frozen phonon method as implemented in the
{\sc phonopy} software package.\cite{parl97,togo08} After the normal
mode coordinates were identified, total energy calculations were
performed as a function of the $\qlz$, $\qlx$, and $\qhx$ coordinates
for values ranging between $-3$ and $3$ \ampunit\ with a step of 0.1
\ampunit\ to obtain the energy surfaces $V(\qlz,\qlx,\qhx)$. These
were then fitted to polynomials given in Eq.~\ref{eq:full} to obtain
normal mode anharmonicities and nonlinear couplings between the three
coordinates. For the materials that I explored, polynomials with
anharmonicities up to twentieth order and nonlinearities up to eighth
order were needed to ensure accurate fit to the calculated energy
surfaces. Since the polynomial fits the calculated energy surfaces
almost exactly, there are no approximations in the calculations of the
nonlinear couplings, beyond that for the exchange-correlation
functional.

The Born effective charges were calculated using density functional
perturbation theory,\cite{dfpt} and a larger $16\times16\times16$
$k$-point grid was used in these calculations. The calculated Born
effective charges and phonon mode eigenvectors were used to calculate
the mode effective charge $\zmstar$ that determines the strength of
the coupling of light to the pumped phonon mode from first
principles.\cite{gonz97} The coupled equations of motion for the three
coordinates were numerically solved using the {\sc lsode} subroutine
of the {\sc octave} software package.\cite{octave}

\subsection{Identifying light-induced ferroelectricity}

Phase transitions cannot occur at short timescales in nonequilibrium
conditions, and any light-induced ferroelectricity will disappear once
the external light source vanishes. Therefore, it is necessary to
establish an unequivocal protocol for identifying light-induced
ferroelectricity. Examining the intensity and phase of the second
harmonic generation of the transmitted probe pulse is a convenient way
to study ferroelectricity in pump-probe
experiments,\cite{taka06,talb08} and it will be necessary to
distinguish between light-induced ferroelectricity and a
long--time-period excitation that both generate second harmonics if
the probe pulse is shorter than the period of the low-frequency mode.
For the purpose of this discussion, a light-induced ferroelectric
state is deemed to have occurred both if the phase of the second
harmonics does not change and the intensity of the second harmonics
shows at least two peaks over the full width at half maximum (FWHM)
duration of the pump pulse. Therefore, the pump pulse duration should
in general be larger than the period of the equilibrium-condition
lowest frequency mode to establish light-induced
ferroelectricity. However, this is not a strict condition, and other
well-defined criteria could also be specified. In particular, the
lowest frequency oscillations could (and indeed does) occur with a
larger frequency in the rectified state, and any method (such as time
resolved x-ray diffraction) that can distinguish oscillations about a
displaced position can establish light-induced ferroelectricity.

\section{Results and Discussions}

\begin{figure}
  \includegraphics[width=\columnwidth]{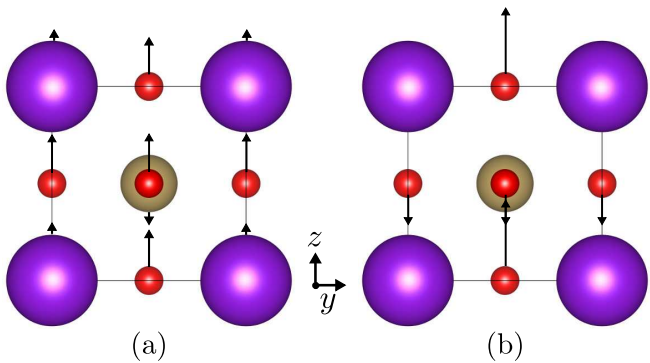}
  \caption{(Color online) Displacement patterns of the (a) lowest
    frequency $\qlz$ and (b) highest frequency $\qhz$ modes of the
    cubic phase of KTaO$_3$. The $x$ and $y$ components of these
    triply degenerate modes can be obtained by appropriate
    rotation. The big, medium, and small spheres denote K, Ta, and O,
    respectively. }
  \label{fig:disp}
\end{figure}

\subsection{Cubic KTaO$_3$}
The paraelectric phase of several $AB$O$_3$ perovskite oxides occurs
in the cubic structure. So it is natural to ask if ferroelectricity
can be induced in these cubic paraelectrics by a midinfrared
excitation of their infrared-active phonon modes. These materials have
five atoms per unit cell, and they thus have four triply degenerate
optical phonon modes at the zone center. Factor group analysis shows
that three of these modes have the irreducible representation
$T_{1u}$, and these are infrared active. The remaining one has the
irreducible representation $T_{2u}$ and is optically inactive.
Ferroelectricity is generally ascribed to a dynamical instability of
an infrared-active transverse optic phonon mode. Indeed, most
ferroelectric materials show a characteristic softening of an infrared
transverse optic mode as the transition temperature is
approached.\cite{scot74} Here I investigate if a similar softening and
instability of the lowest frequency $\Tou$ mode can be achieved by an
intense laser-induced excitation of the highest frequency $\Tou$ mode
in the representative case of cubic KTaO$_3$.

The calculated phonon frequencies of cubic KTaO$_3$ using the relaxed
PBEsol lattice parameter of 3.99 \AA\ are $\oml =$ 85 cm$^{-1}$ and
$\omh =$ 533 cm$^{-1}$ for the lowest and highest frequency $\Tou$
modes, respectively. These are in good agreement with previously
calculated values.\cite{sing96} They also agree well with the
frequencies obtained from hyper-Raman scattering experiments at room
temperature.\cite{vogt84} The atomic displacement patterns due to
these two modes are shown in Fig.~\ref{fig:disp}. Without loss of
generality, I consider the case where the $x$ component of the highest
frequency $\Tou$ mode $\qhx$ is pumped by an intense light source and
study how such an excitation changes the dynamics of the lowest
frequency $\Tou$ mode along the longitudinal $\qlx$ and transverse
$\qlz$ coordinates. I ignore the dynamics along the second transverse
coordinate $\qly$ as its dynamics will be qualitatively similar to
that of the $\qlz$ coordinate.

\begin{figure}
  \includegraphics[width=\columnwidth]{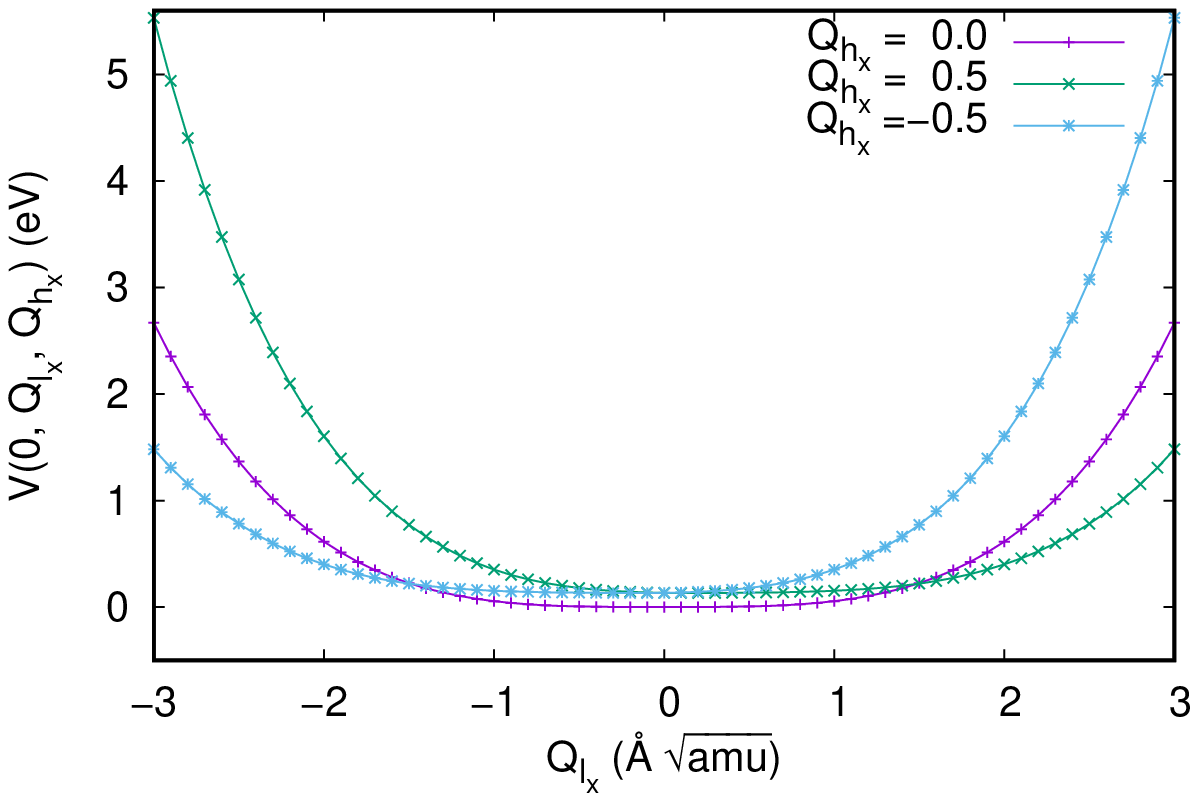}
  \caption{(Color online) Total energy as a function of the
    longitudinal $\qlx$ coordinate for several values of the $\qhx$
    coordinate for cubic KTaO$_3$.}
  \label{fig:cubic-qlx-qhx}
\end{figure}

\subsubsection{Dynamics of the lowest frequency longitudinal component}

Fig.~\ref{fig:cubic-qlx-qhx} shows several total energy $V(\qlz, \qlx,
\qhx)$ curves along the projection $\qlz = 0$.  The curves are not
symmetric upon reflection at $\qlx = 0$ and the $+\qhx$ and $-\qhx$
curves do not overlap. This indicates the presence of coupling terms
that have odd orders of $\qlx$ and $\qhx$. A polynomial fit of the
energy surface shows that the coupling terms $t_1 \qlx^3 \qhx$, $t_2
\qlx^2 \qhx^2$ and $t_3 \qlx \qhx^3$ are all large relative to the
harmonic term $\oml^2$ of the lowest frequency mode (see
Table~\ref{tab:coup}).  The presence of these couplings is consistent
with the symmetry requirements. Since the equilibrium structure has
inversion symmetry and we are considering two odd modes along the same
direction, any term $\qlx^m \qhx^n$ is allowed as long as $m + n =
even$. The next allowed order of coupling is $\qlx^m \qhx^n$ with $m+n
= 6$. These are an order of magnitude smaller than the $m+n = 4$ terms
(see Table~\ref{tab:coup-app} in the Appendix), but they are
comparable in magnitude to the harmonic term $\oml^2$.

\begin{table}
  \caption{\label{tab:coup} The coefficients of the harmonic and
    nonlinear coupling terms of cubic and strained KTaO$_3$. The units
    of a $Q^mQ^n$ term are meV \AA$^{-(m+n)}$
    amu$^{\frac{-(m+n)}{2}}$. The sign of the coupling is relevant
    only when the coordinates come with even powers.}
  \begin{ruledtabular}
    \begin{tabular}{l c d{4.2} d{4.2}}
      coefficient & order & \multicolumn{1}{c}{cubic} & \multicolumn{1}{c}{strained} \\
      \hline
      $\omlz^2$ & $\qlz^2$        &   27.06  &    1.39 \\
      $\omlx^2$ & $\qlx^2$        &   27.06  &   55.27 \\
      $\omh^2$  & $\qhx^2$        & 1043.77  & 1136.10 \\
      $t_1$     & $\qlx^3 \qhx$   & -118.35  &   97.38 \\
      $t_2$     & $\qlx^2 \qhx^2$ &  215.00  &  208.76 \\
      $t_3$     & $\qlx \qhx^3$   & -175.58  &  195.22 \\
      $l$       & $\qlz^2 \qhx^2$ & -5.95    &   -5.81 \\
      $m_1$     & $\qlz^4 \qhx^2$ & -1.03    &   -1.00 \\
      $m_2$     & $\qlz^2 \qhx^4$ & -3.05    &   -4.12 \\
    \end{tabular}
  \end{ruledtabular}
\end{table}

The nonlinear couplings between the $\qlx$ and $\qhx$ modes impart a
force equal to $-\partial V / \partial \qlx $ along the $\qlx$
coordinate. This force is finite and large when the $\qhx$ mode is
externally excited by an intense light source. The lowest order
nonlinear terms of this force are $-\partial V / \partial \qlx = -3
t_1 \qlx^2 \qhx - 2 t_2 \qlx \qhx^2 - t_3 \qhx^3$. The $-t_3\qhx^3$
term acts as a nonresonant oscillating force to the $\qlx$ mode. The
effect of the $-3 t_1 \qlx^2 \qhx$ term would average over the slow
oscillation of the $\qlx$ mode relative to that of the $\qhx$
mode. The $-2 t_2 \qlx \qhx^2$ term affects a time-dependent
modulation of the frequency of the $\qlx$ mode, and it does not cancel
over the slow oscillation of the $\qlx$ mode because $\qhx^2$ has a
nonzero time average.\cite{fors11b, sube14} Unfortunately, the sign of
$t_2$ is positive, so the frequency of the $\qlx$ mode increases as
the $\qhx$ mode is pumped. A similar analysis of the next order
$\qlx^m\qhx^n$ terms with $m+n= 6$ also shows that the $\qlx$ mode
does not soften due to the effects of nonlinear coupling terms.

\subsubsection{Dynamics of the lowest frequency transverse component} 

\begin{figure}[t]
  \includegraphics[width=\columnwidth]{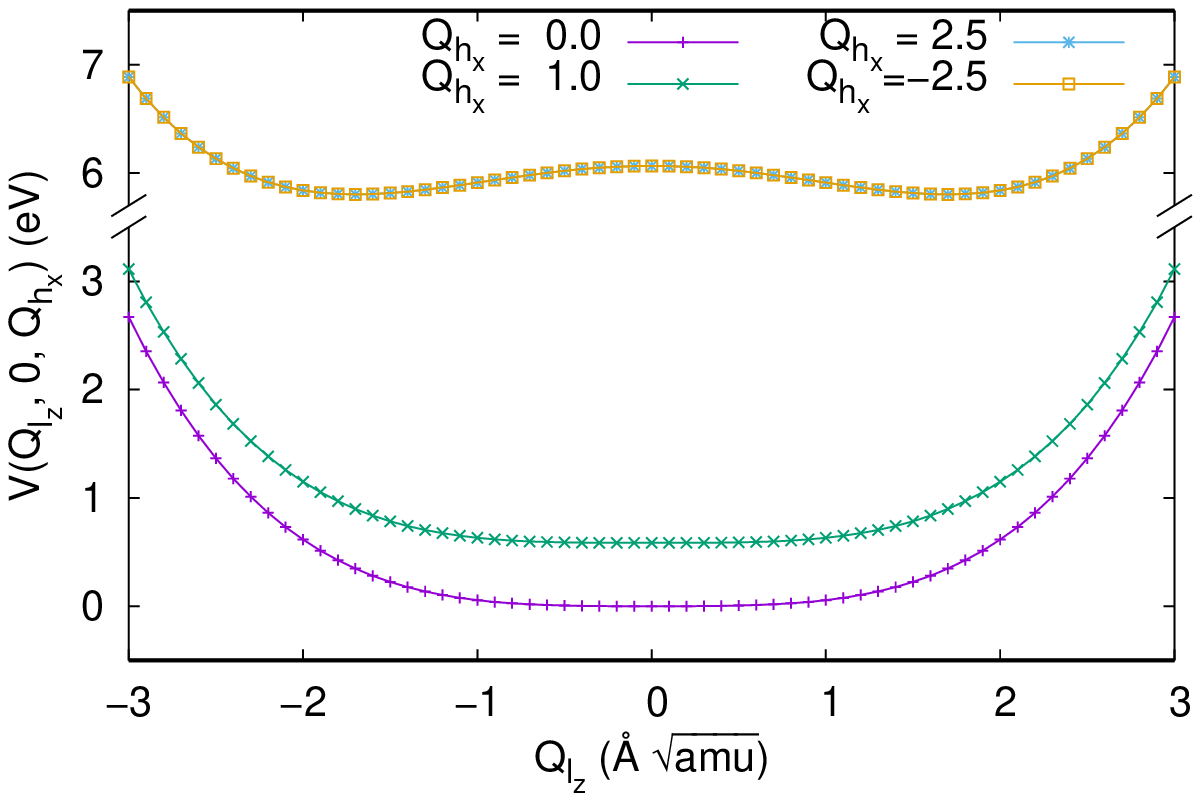}
  \caption{(Color online) Total energy as a function of the transverse
    $\qlz$ coordinate for several values of the $\qhx$ coordinate for
    cubic KTaO$_3$.}
  \label{fig:cubic-qlz-qhx}
\end{figure}

What about the dynamics of the transverse component $\qlz$ of the
lowest frequency mode? Fig.~\ref{fig:cubic-qlz-qhx} shows several
total energy $V(\qlz,\qlx,\qhx)$ curves along the projection
$\qlx=0$. One immediately notices that the curves are symmetric upon
reflection at $\qlz =0$ and that the $-\qhx$ and $\qhx$ curves
overlap, showing that only even powers of both $\qlz$ and $\qhx$
appear in the nonlinear coupling terms. This is again consistent with
the symmetry requirements, which does not allow products with odd
powers of mutually perpendicular components $\qlz$ and $\qhx$. The
coefficients of the lowest order nonlinear terms $l \qlz^2 \qhx^2$,
$m_1 \qlz^4 \qhx^2$, and $m_2 \qlz^2 \qhx^4$ are given in
Table~\ref{tab:coup}. They are all at least twenty times smaller than
the magnitude of the quartic order couplings between the $\qlx$ and
$\qhx$ coordinates. Nevertheless, the sign of the coupling
coefficients between the $\qlz$ and $\qhx$ modes are such that these
terms soften the frequency of the $\qlz$ mode, as one sees by
analyzing the forcing terms due to these nonlinear couplings
$-\partial V / \partial \qlz = - 2 l \qlz \qhx^2 - 4 m_1 \qlz^3 \qhx^2
- 2 m_2 \qlz \qhx^4$. Each term in the previous expression has even
powers of the $\qhx$ coordinate, which ensures that their effects are
not averaged over the slow oscillation of the $\qlz$
mode. Furthermore, all these terms are proportional to odd powers of
the $\qlz$ coordinate, which causes the frequency of the $\qlz$ mode
to change as $\omlz^2 \rightarrow \omlz^2 [1 + (2 l \qhx^2\!(t) + 4
  m_1 \qlz^2\!(t) \qhx^2\!(t) + 2 m_2 \qhx^4\!(t)) / \omlz^2 ]$. Since
the coupling constants are negative, this should lead to a softening
of the transverse $\qlz$ mode when the $\qhx$ mode is externally
pumped.

The above discussion is not sufficient to convincingly argue that the
transverse $\qlz$ mode will become dynamically unstable when the
$\qhx$ mode is externally pumped. There are two counteracting
processes that may preclude this from happening. First, the coupling
between the $\qlx$ and $\qhx$ modes is at least twenty times larger.
As a result, the $\qlz$ component may receive a much smaller
proportion of the external force due to nonlinear coupling with the
$\qhx$ mode that is not sufficient to make this mode dynamically
unstable when the latter is pumped. Moreover, I find that the coupling
term $p\qlz^2\qlx^2$ to be positive and larger than the term
$l\qlz^2\qhx^2$ (see Table~\ref{tab:coup-app} in the Appendix). When
the $\qlx$ component oscillates with a large amplitude, this will
provide an additive factor that increases the frequency of the $\qlz$
component.

To settle this issue, I numerically solved the coupled equations of
motion of the three coordinates $\qlz$, $\qlx$, and $\qhx$, which are
\begin{eqnarray}
  \label{eq:cubic}
  \ddot{Q}_{\textrm{h$_x$}} + \gamma_{\textrm{h}} \dot{Q}_{\textrm{h$_x$}} + \omh^2 \qhx &
  = &  -\frac{\partial V^{\textrm{nh}}(\qlz,\qlx,\qhx)}{\partial \qhx}
  + F(t) \nonumber \\
  \ddot{Q}_{\textrm{l$_x$}} + \gamma_{\textrm{l}} \dot{Q}_{\textrm{l$_x$}} + \oml^2 \qlx &
  = &  -\frac{\partial V^{\textrm{nh}}(\qlz,\qlx,\qhx)}{\partial \qlx}
  \nonumber \\
  \ddot{Q}_{\textrm{l$_z$}} + \gamma_{\textrm{l}} \dot{Q}_{\textrm{l$_z$}} + \oml^2 \qlz &
  = &  -\frac{\partial V^{\textrm{nh}}(\qlz,\qlx,\qhx)}{\partial \qlz}.
\end{eqnarray}
Here, $V^{\textrm{nh}}(\qlz,\qlx,\qhx)$ is the nonharmonic part of the
polynomial that fits the calculated energy surface, and it includes
both the anharmonicities of each coordinates as well as the nonlinear
couplings between these coordinates. The full expression for
$V^{\textrm{nh}}(\qlz,\qlx,\qhx)$ is given in Eq.~\ref{eq:full} in the
Appendix. In addition to the numerically large nonlinear couplings
discussed above, it includes anharmonicities up to the sixteenth order
and nonlinear couplings up to the eighth order. $\gamma_{\textrm{h}}$
and $\gamma_{\textrm{l}}$ are the damping coefficients of the highest
and lowest frequency $\Tou$ modes, respectively. They are taken to be
ten percent of the respective harmonic terms. $F(t) = \zmstarhx E_0
\sin (\Omega t) e^{-t^2/2(\sigma/2\sqrt{2\ln2})^2}$ is the external
force experienced by the $\qhx$ coordinate due to a light pulse of peak
electric field $E_0$. The calculated mode effective charge of the
$\qhx$ mode of cubic KTaO$_3$ is $\zmstarhx = -1.07e$
amu$^{-\frac{1}{2}}$. A pump with a frequency of $\Omega = 1.01
\Omega_{\textrm{h}}$ and FWHM of $\sigma = 2.0$ ps has been used. The
use of a long pulse width is just to illustrate many oscillation
cycles. The physics does not change when I use a pulse duration larger
than $1/\oml$.

\begin{figure}
  \includegraphics[width=\columnwidth]{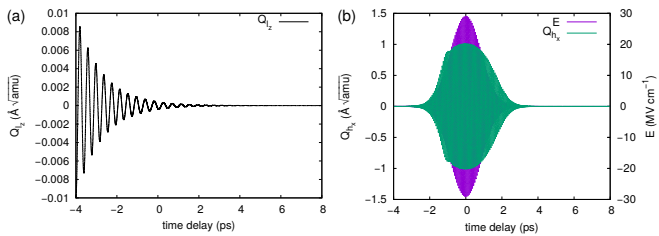}
  \caption{(Color online) Dynamics of the (a) $\qlz$ and (b) $\qhx$
    coordinates of cubic KTaO$_3$ after the $\qhx$ coordinate is
    pumped by a pump pulse $E$ with FWHM of 2 ps as shown in (b).}
  \label{fig:cubic-dyn}
\end{figure}

The result of the numerical integration of Eq.~\ref{eq:cubic} in a
highly nonlinear regime is shown in Fig.~\ref{fig:cubic-dyn}. This was
obtained with a large peak electric field of $E_0 = 30$ MV cm$^{-1}$
that caused the pumped $\qhx$ mode to oscillate with a maximum
amplitude of 1.1 \ampunit\ [Fig.~\ref{fig:cubic-dyn}(b)].  In this
regime, the $\qlx$ mode oscillates about the equilibrium position with
a maximum amplitude of 0.25 \ampunit\ (not shown). The frequency of
its transverse counterpart $\qlz$ does soften by around $\sim$5\%.
But relatively little force is imparted to the $\qlz$ component, and
its oscillations about the equilibrium position are damped even during
the duration of the pump pulse [Fig.~\ref{fig:cubic-dyn}(a)].  I
performed similar calculations for pump fields up to 100 MV cm$^{-1}$
but was not able to find any instances where the $\qlz$ mode becomes
dynamically unstable.

These calculations show that a dynamical instability of the lowest
frequency infrared mode of cubic KTaO$_3$ cannot be achieved by a
midinfrared excitation of its highest frequency infrared
mode. However, this does not allow us to infer that light-induced
dynamical instability cannot occur in any cubic paraelectric. Indeed,
if I artificially increase the coefficient of the $l\qlz^2\qhx^2$ term
by six times, I am able to obtain a solution where the $\qlz$
coordinate oscillates about a displaced position during the duration
of the pump pulse. Such a large coupling between the pumped
high-frequency mode and the transverse component of the low-frequency
mode may exist in some materials.

\subsection{Strained KTaO$_3$}
If the coupling between the externally pumped highest frequency $\Tou$
mode and the component of the lowest frequency $\Tou$ mode
longitudinal to the pumped mode could be weakened in cubic KTaO$_3$,
the transverse component of the lowest frequency mode would develop a
light-induced dynamical instability. An effective way of achieving
this is by raising the frequency of the longitudinal component
relative to that of the transverse component. This can be accomplished
by applying a biaxial strain on KTaO$_3$ via an epitaxial growth on an
appropriate substrate.

I performed calculations on KTaO$_3$ with 0.6\% compressive biaxial
strain. This can be achieved, for example, by growing KTaO$_3$ on a
GdScO$_3$ substrate. The calculated PBEsol lattice parameters of thus
strained KTaO$_3$ are $a=b= 3.965$ and $c = 4.0$ \AA. Upon the
application of a biaxial strain, the $\Tou$ mode of the cubic phase
splits into a nondegenerate $A_{2u}$ mode and a doubly degenerate
$E_u$ mode. The $A_{2u}$ phonons involve atomic motions along the $z$
axis while the atoms move in the $xy$ plane for the $E_u$ phonons. The
calculated values for the lowest frequency $A_{2u}$ and $E_u$ modes
are $\omlz = 20$ and $\omlx = \omly = 122$ cm$^{-1}$,
respectively. The highest frequency $E_u$ phonon that should be
externally pumped has a frequency of $\omh = 556$ cm$^{-1}$.

To find out whether the lowest frequency $A_{2u}$ mode $\qlz$ of
strained KTaO$_3$ develops a dynamical instability when the $x$
component of the highest frequency $E_u$ mode $\qhx$ is intensely
excited by a light source, I again started my investigation by
calculating the total energy surface $V(\qlz,\qlx,\qhx)$ as a function
of the three coordinates using density functional calculations. The
nonlinear couplings between the $\qlz$, $\qlx$, and $\qhx$ coordinates
of strained KTaO$_3$ have the same symmetry requirements as discussed
for the cubic case, and a fit of a general polynomial to the
calculated first-principles $V(\qlz,\qlx,\qhx)$ shows that same orders
of nonlinearities are present in both the cases. As a comparison of
the numbers presented in Table~\ref{tab:coup} shows (see also
Table~\ref{tab:coup-app} in the Appendix), the nonlinear couplings in
the two cases do not differ by a large amount. The crucial difference
between the two cases is that the frequencies of the $\qlz$ and $\qlx$
coordinates are different in the strained case ($\omlz = 20$ and
$\omlx = 122$ cm$^{-1}$), whereas they are equal in the cubic case
($\omlz = \omlx = 85$ cm$^{-1}$). This has a profound effect in the
dynamics of the $\qlz$ coordinate because the forces experienced by a
coordinate due to the nonlinear couplings are weighted by the square
of the frequency of the coordinate. We can see that
$\frac{1}{20^2}\frac{\partial V}{\partial \qlz}$ is likely to be much
larger than $\frac{1}{122^2}\frac{\partial V}{\partial \qlx}$ or
$\frac{1}{85^2}\frac{\partial V}{\partial \qlx}$. In simple words, the
$\qlz$ coordinate of strained KTaO$_3$ gets much larger proportion of
the force than the $\qlz$ coordinate of cubic KTaO$_3$ because the
frequency of the $\qlz$ mode is much smaller in the strained structure
compared to the cubic structure. Is this change big enough to result
in a light-induced dynamical instability of the $\qlz$ mode in
strained KTaO$_3$?

I again solved the coupled equations of motion of the three
coordinates $\qhx$, $\qlx$, and $\qlz$ as given by Eq.~\ref{eq:cubic}
for the case of strained KTaO$_3$. This time I used the potential
$V^{\textrm{nh}}(\qlz,\qlx,\qhx)$ obtained for strained KTaO$_3$ from
first principles. The polynomial expression used in the calculations
and the numerical values of the coefficients for all the terms in the
polynomial that fit the calculated energy surface are given in the
Appendix. A pump pulse with an FWHM of 2 ps ($ > 1/\omlz$) and
frequency $1.01\omh$ is again used to excite the $\qhx$ mode. The mode
effective charge of the $\qhx$ mode in the strained structure is
$\zmstarhx = -1.15e$ amu$^{-\frac{1}{2}}$.

\begin{figure}
  \includegraphics[width=\columnwidth]{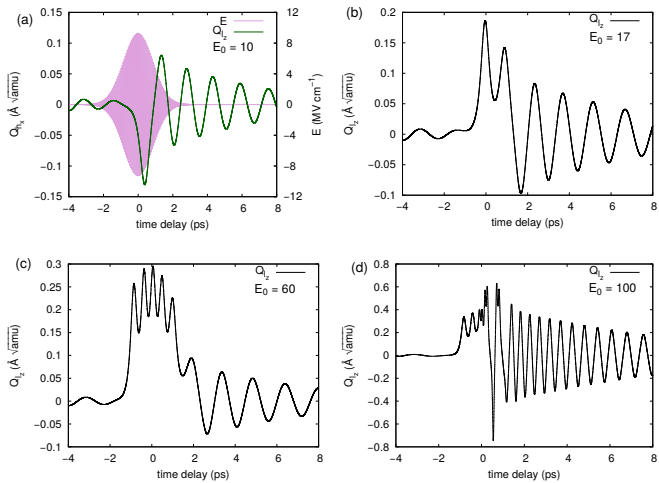}
  \caption{(Color online) Dynamics of the $\qlz$ coordinate of
    strained KTaO$_3$ after the $\qhx$ coordinate is pumped by a pump
    pulse $E$ with FWHM of 2 ps. The dynamics for four different
    values of the peak electric field $E_0$ (MV cm$^{-1}$) of the pump
    pulse are shown.}
  \label{fig:tetra-dyn}
\end{figure}

Fig.~\ref{fig:tetra-dyn} shows the results of the numerical
integration of these equations for three different regimes of dynamics
of the $\qlz$ coordinate. At relatively small peak electric fields of
the pump ($E_0 < 1$ MV cm$^{-1}$), the $\qlz$ mode oscillates about
the equilibrium position with its harmonic frequency (not shown). As
the peak electric field is increased, the frequency of the $\qlz$ mode
decreases during the duration that the $\qhx$ mode is being pumped
[Fig.~\ref{fig:tetra-dyn}(a)]. As discussed above, this is due to the
negative values of the coefficients of the coupling terms $l \qlz^2
\qhx^2$, $m_1 \qlz^4 \qhx^2$, and $m_2 \qlz^2 \qhx^4$ that cause a
light-induced softening of the $\qlz$ coordinate.  Since the duration
of the pump pulse is finite, I naturally do not observe the period of
the $\qlz$ mode diverge.  Instead, beyond a critical value of the peak
electric field of the pump ($E_0^c \simeq 17$ MV cm$^{-1}$for $\sigma
= 2$ ps), the $\qlz$ coordinate oscillates about a displaced position
and has a non-zero value while the $\qhx$ mode is being pumped
[Figs.~\ref{fig:tetra-dyn}(b, c)]. In this rectified regime, the
average potential felt by the $\qhz$ mode has a double-well structure,
and this mode is oscillating about one of the minima.  The
displacement along the $\qlz$ coordinate is also amplified strongly in
this regime. Since the $\qlz$ mode is infrared active, this implies
that the material is in a broken symmetry state with a finite dipole
moment while the $\qhx$ mode is externally pumped.

The frequency of the $\qlz$ oscillations in the rectified state
increases as the peak electric field of the pump is increased beyond
the critical threshold. This is evident from a comparison of
Figs.~\ref{fig:tetra-dyn}(b) and (c), which shows that the frequency
of the $\qlz$ mode doubles as the peak electric field $E_0$ is
increased from 17 to 60 MV cm$^{-1}$. This increase occurs because the
double-well potential for the $\qlz$ coordinate becomes deeper as the
amplitude of the $\qhx$ oscillations increases.

When the peak electric field is increased further ($E_0 > 75$ MV
cm$^{-1}$), the $\qlz$ mode oscillates with a large amplitude and high
frequency about the equilibrium position
[Fig.~\ref{fig:tetra-dyn}(d)]. In this regime, the kinetic energy
imparted to the $\qlz$ mode is larger than the depth of the double
wells. As a result, the oscillation of the $\qlz$ mode stops being
confined to one of the double wells, and the rectified behavior along
the $\qlz$ coordinate is no longer observed. Even though the
light-induced broken-symmetry phase is stabilized only for a range of
values of the peak electric field of the pump, this range $17 < E_0 <
75$ MV cm$^{-1}$ is both wide and approachable enough to make the
light-induced ferroelectric state experimentally accessible.

The existence of a critical threshold above which the $\qlz$
coordinate is rectified and the presence of three different regimes
for the dynamics of this coordinate is consistent with the analysis of
a $Q_1^2Q_2^2$ nonlinear coupling between two different normal mode
coordinates as presented in Ref.~\onlinecite{sube14}. These features
should be present in the experiments to confirm the predictions
made in this work. The critical pump amplitude depends on the
frequencies of the $\qlz$ and $\qhx$ modes and the coupling
coefficient, as well as the pump pulse length and the initial
condition (i.e. the r.m.s.\ displacement of the $\qlz$ mode at a
particular temperature).\cite{sube14} For a pump pulse with FWHM of 2
ps, I find that the $\qlz$ mode starts to get rectified when the peak
electric field is $E_0$ = 17 MV cm$^{-1}$. With this pump pulse, the
$\qhx$ mode is oscillating with an amplitude of 0.9 \ampunit, which
corresponds to a maximum change in the Ta-apical O bond length of 0.2
\AA\ (that is, 10\%). The $\qhx$ mode oscillates with an amplitude of
1.3 \ampunit\ when the peak electric field is $E_0 = 75$ MV
cm$^{-1}$. This is a modest increase in the energy of the $\qhx$ mode
due to the pump, and it indicates that a significant fraction of the
pumped energy goes to maintaining the rectified state along the $\qlz$
coordinate.  However, the light-induced ferroelectric displacement
along the $\qlz$ coordinate is quite small because of the small
magnitude of the couplings between $\qlz$ and $\qhx$ modes. The
average displacement along $\qlz$ is $\sim$0.1 and $\sim$0.2
\ampunit\ for $E_0 = 17$ and 75 MV cm$^{-1}$, respectively, which
results in the change of Ta-apical O distance by 0.015--0.030 \AA.

Curiously, the critical pump threshold obtained for strained KTaO$_3$
is noticeably smaller than what is expected for a $\qlz^2\qhx^2$
coupling. In the total energy calculations, the $\qlz$ mode starts
developing instability when $\qhx$ is above 0.7 \ampunit. So the
critical $\qhx$ amplitude should be $0.7\sqrt{2} = 1.0$
\ampunit.\cite{sube14} Instead, I find that the $\qlz$ mode becomes
unstable when the $\qhx$ amplitude is $0.9$ \ampunit. This reduction
in the critical threshold is due to the presence of a large and
negative sixth order coupling terms $m_1\qlz^4\qhx^2$ and
$m_2\qlz^2\qhx^4$. Both these terms give a subtractive contribution to
the effective, light-induced frequency of the $\qlz$ mode, which
hastens its instability as a function of the pump intensity.

\subsection{Abruptly halting light-induced ferroelectricity} 

For light-induced ferroelectricity to be useful in applications, it is
necessary to be able to control the light-induced phase at will in an
all-optical setup. In this context, this means having the capability
to switch off the rectification of the $\qlz$ mode while the $\qhx$
mode is being pumped. The quartic order odd $\qlz^3\qhz$ and
$\qlz\qhz^3$ couplings in the longitudinal direction can be used to
our advantage for this purpose. To investigate this possibility, I
consider an experiment where an overlapping pulse polarized along
$\qhz$ comes at an arbitrary delay with respect to the
rectification-causing pulse that pumps the $\qhx$ mode. I study the
resulting dynamics along the $\qlz$ coordinate by solving the coupled
equations of motion for the four coordinates ($\qlz,\qlx,\qhx,$ and
$\qhz$). The equations of motions are obtained from the potential
$V^{\textrm{nh}}(\qlz,\qlx,\qhx) + V^{\textrm{nh}}(\qlz,\qhz)$. For
computational efficiency, I do not consider the full potential
$V^{\textrm{nh}}(\qlz,\qlx,\qhx,\qhx)$ spanned by the four
coordinates.

\begin{figure}
  \includegraphics[width=\columnwidth]{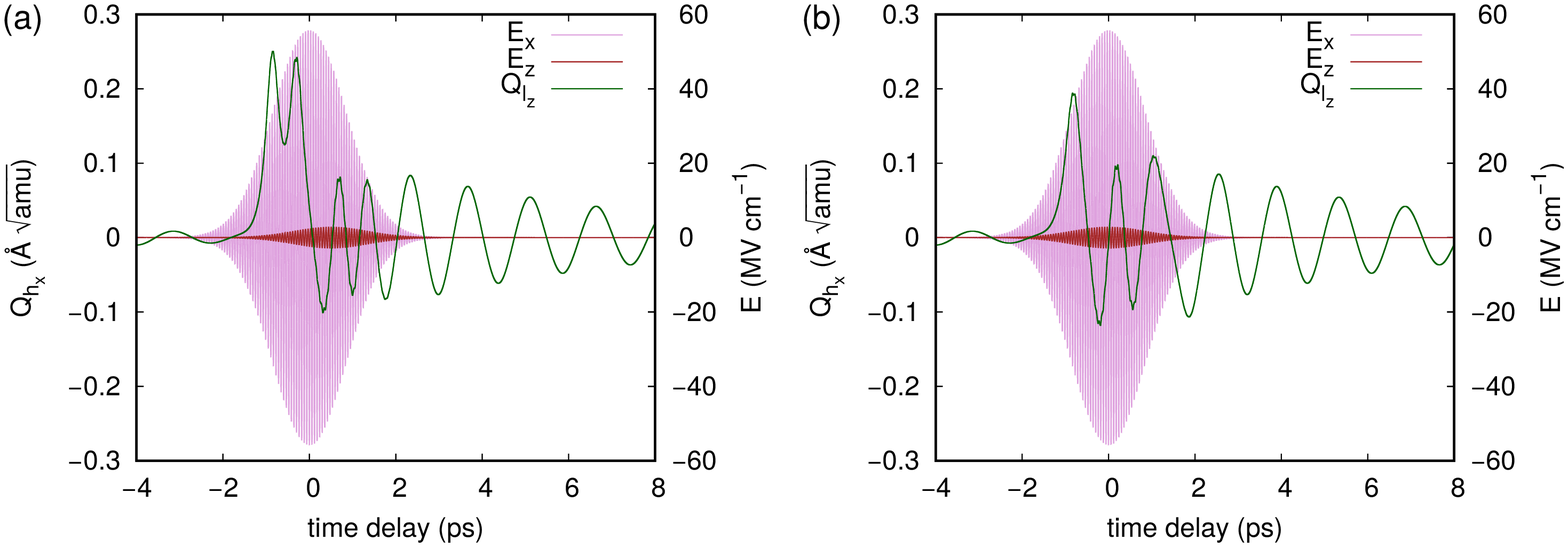}
  \caption{(Color online) Dynamics of the $\qlz$ coordinate of
    strained KTaO$_3$ for the delays of (a) 0.5 and (b) 0.0 ps between
    the $E_{\textrm{h$_x$}}$ and $E_{\textrm{h$_z$}}$ pulses that pump
    $\qhx$ and $\qhz$ modes, respectively.}
  \label{fig:rect-cont}
\end{figure}

The results for the delays of 0.5 and 0.0 ps between the pump pulses
$E_{\textrm{h$_x$}}$ and $E_{\textrm{h$_z$}}$ that excite the $\qhx$
and $\qhz$ coordinates, respectively, are shown in
Fig.~\ref{fig:rect-cont}. The peak electric fields of
$E_{\textrm{h$_x$}}$ and $E_{\textrm{h$_z$}}$ are 60 and 3 MV
cm$^{-1}$, respectively, and their FWHM is 2 ps. The pump frequencies
are 1.01 times the respective phonon frequencies and the mode
effective charge of $\qhz$ is $\zmstarhz = -1.05e$
amu$^{-\frac{1}{2}}$. Note that excitation by only
$E_{\textrm{h$_x$}}$ causes rectification of the $\qlz$ mode during
the FWHM of the pulse [Fig.~\ref{fig:tetra-dyn}(c)]. However, an
overlapping excitation by another weak pulse $E_{\textrm{h$_z$}}$
immediately suppresses the light-induced rectification of the $\qlz$
mode. Even a weak longitudinal pump is efficient in halting the
rectification because the quartic order odd couplings between $\qlz$
and $\qhz$ modes are much larger than the couplings in the transverse
direction.

\section{Summary and Conclusions}

In summary, I have shown that midinfrared pulses can be used to
stabilize nonequilibrium ferroelectricity in strained KTaO$_3$, which
is paraelectric at equilibrium conditions. This phenomenon relies on a
quartic $l\qlz^2\qhx^2$ coupling between the highest frequency
infrared-active phonon mode $\qhx$ and the lowest frequency
infared-active mode $\qlz$ that is transverse to $\qhx$. Density
functional calculations show that the coupling constant $l$ is
negative, which causes the $\qlz$ mode to soften when the $\qhx$ mode
is externally pumped. The rectification along the $\qlz$ coordinate
occurs only above a critical electric field of the pump pulse,
demonstrating that this light-induced symmetry breaking is a unique
nonperturbative effect. Such a threshold behavior should be observed
in experiments to corroborate the predictions made in this paper.
Additionally, the $\qlz^4\qhx^2$ and $\qlz^2\qhx^4$ couplings are
large, and this makes the rectified regime more accessible. A first
principles calculation of the coupling between light and the $\qhx$
mode shows that ferroelectricity can be induced in strained KTaO$_3$
by a midinfrared pulse with a peak electric field of 17 MV cm$^{-1}$
and a duration of 2 ps. Furthermore, large odd quartic couplings
$\qlz^3\qhz$ and $\qlz\qhz^3$ between $\qlz$ and the highest frequency
infrared-active mode $\qhz$ longitudinal to $\qlz$ makes it possible
to arbitrarily switch off the induced ferroelectricity by pumping the
$\qhz$ mode with another weak pulse. I find that similar nonlinear
interactions exist in SrTiO$_3$ and LaAlO$_3$, and this technique
could be generally applied to other transition metal oxide
paraelectrics.

At a more basic level, I have shown that materials can exhibit various
nonlinear interactions between different dynamical degrees of freedom
that have hitherto been overlooked. These interactions enable us to
induce and control broken-symmetry phases using light, whose
oscillating electric and magnetic fields average to zero by
definition. Furthermore, I have demonstrated that the nonlinear
interactions can be effectively modified by applying strain. This
motivates experiments that combine the disparate fields of nonlinear
optics and heterostructuring to achieve materials control in an
interesting manner. In a broader perspective, these nonlinear
interactions may also be present in other classes of systems, and they
might allow us to influence the dynamics of these systems in an
unusual way.

\acknowledgments 

I am grateful to Indranil Paul for helpful discussions. This work was
supported by the European Research Council grants ERC-319286 QMAC and
ERC-61719 CORRELMAT and the Swiss National Supercomputing Center
(CSCS) under project s575.

\appendix

\section{Expressions for total energy surfaces}

Only low order nonlinear couplings that are relatively large were
discussed in the main text. However, if only low order couplings and
anharmonicities are considered, the fit to the calculated total-energy
surfaces are not satisfactory. The dynamics of the coordinates with
and without using the high order couplings also show large
differences, especially at the nonlinear regime. Since the use of the
full polynomial expression in the solutions of the equations of motion
are not computationally demanding, all the numerical results discussed
in this paper were obtained using the full expression given below.

For cubic KTaO$_3$, the following polynomial $V(\qlz,\qlx,\qhx)$
accurately fits the calculated total energy surface spanned by the
three coordinates for values between
$-3.0$ and $3.0$ \ampunit.
\begin{eqnarray}
  \label{eq:full}
  V & = & \frac{1}{2} \oml^2 \qlz^2 + a_4 \qlz^4 + a_6
  \qlz^6 + a_8 \qlz^8 + a_{10} \qlz^{10} \nonumber \\
  & & + a_{12} \qlz^{12} + a_{14} \qlz^{14} + a_{16} \qlz^{16}
  + \frac{1}{2} \oml^2 \qlx^2 + a_4 \qlx^4 \nonumber \\
  & & + a_6 \qlx^6 + a_8 \qlx^8 + a_{10} \qlx^{10} + a_{12} \qlx^{12} + a_{14} \qlx^{14} 
  \nonumber \\
  & & + a_{16} \qlx^{16} + \frac{1}{2} \omh^2 \qhx^2 + c_4 \qhx^4 +
  c_6 \qhx^6 + c_8 \qhx^8 
  \nonumber \\
  & & + c_{10} \qhx^{10} + c_{12} \qhx^{12} + l \qlz^2 \qhx^2 + m_1 \qlz^4 \qhx^2
  \nonumber \\
  & & + m_2 \qlz^2 \qhx^4  + n_1 \qlz^4 \qhx^4 + n_2 \qlz^6 \qhx^2 + n_3 \qlz^2 \qhx^6 \nonumber \\
  & & + t_1 \qlx^3 \qhx + t_2 \qlx^2 \qhx^2 
  + t_3 \qlx \qhx^3 + u_1 \qlx^5 \qhx \nonumber \\
  & & + u_2 \qlx^4 \qhx^2 + u_3 \qlx^3 \qhx^3 
  + u_4 \qlx^2 \qhx^4 + u_5 \qlx \qhx^5 \nonumber \\
  & & + p \qlz^2 \qlx^2 + q_1 \qlz^4 \qlx^2 
  + q_2 \qlz^2 \qlx^4 + r_1 \qlz^4 \qlx^4 
  \nonumber \\
  & & + r_2 \qlz^6 \qlx^2 + r_3 \qlz^2 \qlx^6 
  + d \qlz^2 \qlx \qhx \nonumber \\
  & & + e_1 \qlz^2 \qlx^3 \qhx + e_2 \qlz^2 \qlx^2 \qhx^2 + e_3 \qlz^2 \qlx \qhx^3 
  \nonumber \\
  & & + f_1 \qlz^2 \qlx^5 \qhx + f_2 \qlz^2 \qlx^4 \qhx^2 + f_3 \qlz^2
  \qlx^3 \qhx^3 \nonumber \\
  & & + f_4 \qlz^2 \qlx^2 \qhx^4 + f_5 \qlz^2  \qlx \qhx^5 + g \qlz^4 \qlx \qhx 
  \nonumber \\
  & & + h_1 \qlz^4 \qlx^3 \qhx  + h_2 \qlz^4 \qlx^2 \qhx^2 + h_3 \qlz^4
  \qlx \qhx^3. \nonumber\\
\end{eqnarray}
For strained KTaO$_3$, the potential has additional $a_{18} \qlz^{18}$
and $a_{20} \qlz^{20}$ terms. Also, the coefficients of the $\qlz^n$
and $\qlx^n$ terms are different in the strained case. The
coefficients of the $\qlx^n$ terms for the strained case are denoted
by $b_n$ in Table~\ref{tab:coup-app}. 

The polynomial $V(\qlz,\qhz)$ that fits
the energy surface spanned by the $\qlz$ and $\qhz$ coordinates is
given by
\begin{eqnarray}
  \label{eq:qlzqhz}
  V & = & \frac{1}{2} \omlz^2 \qlz^2 + a_4 \qlz^4 + a_6
  \qlz^6 + a_8 \qlz^8 + a_{10} \qlz^{10} \nonumber \\
  & & + a_{12} \qlz^{12} + a_{14} \qlz^{14} + a_{16} \qlz^{16} + a_{18} \qlz^{18} +
  a_{20} \qlz^{20} \nonumber \\
  & & + \frac{1}{2} \omhz^2 \qhz^2 + d_4 \qhz^4 + d_6 \qhz^6 + d_8 \qhz^8 +
  d_{10} \qhz^{10} \nonumber \\
  & & + d_{12} \qhz^{12} + v_1 \qlz^3 \qhz 
  + v_2 \qlz^2 \qhz^2 + v_3 \qlz \qhz^3 \nonumber \\
  & & + w_1 \qlz^5 \qhz + w_2 \qlz^4 \qhz^2 
  + w_3 \qlz^3 \qhz^3 + w_4 \qlz^2 \qhz^4  \nonumber \\
  & & + w_5 \qlz \qhz^5. 
\end{eqnarray}

\begin{table*}
  \caption{\label{tab:coup-app} The coefficients of the harmonic,
    anharmonic and nonlinear coupling terms of cubic and strained
    KTaO$_3$. The units of a $Q^mQ^nQ^p$ term are meV \AA$^{-(m+n+p)}$
    amu$^{\frac{-(m+n+p)}{2}}$. The sign of the coupling is relevant
    only when the coordinates come with even powers.}
  \begin{ruledtabular}
    \begin{tabular}{l c d{4.2} d{4.7} l c d{4.7}}
      coefficient & order & \multicolumn{1}{c}{cubic} & \multicolumn{1}{c}{strained} & coefficient & order & \multicolumn{1}{c}{strained} \\
      \hline
      $\omlz^2$ & $\qlz^2$                &   27.06                   &    1.39              & $\omhz^2$ & $\qhz^2$    & 1034.38              \\
      $\omlx^2$ & $\qlx^2$                &   27.06                   &   55.27              & $b_4$     & $\qlx^4$    &   36.56              \\
      $\omh^2$  & $\qhx^2$                & 1043.77                   & 1136.10              & $b_6$     & $\qlx^6$    &   -5.05              \\
      $a_4$     & $\qlz^4$                &   47.55                   &   51.72              & $b_8$     & $\qlx^8$    &    1.12              \\
      $a_6$     & $\qlz^6$                &   -6.45                   &   -8.69              & $b_{10}$  & $\qlx^{10}$ &   -1.79\times10^{-1} \\
      $a_8$     & $\qlz^8$                &    1.47                   &    2.73              & $b_{12}$  & $\qlx^{12}$ &    1.85\times10^{-2} \\
      $a_{10}$  & $\qlz^{10}$             &   -2.35\times10^{-1}      &   -6.91\times10^{-1} & $b_{14}$  & $\qlx^{14}$ &   -1.07\times10^{-3} \\
      $a_{12}$  & $\qlz^{12}$             &    2.43\times10^{-2}      &    1.28\times10^{-1} & $b_{16}$  & $\qlx^{16}$ &    2.64\times10^{-5} \\
      $a_{14}$  & $\qlz^{14}$             &   -1.41\times10^{-3}      &   -1.61\times10^{-2} & $d_4$    & $\qhz^4$        &   61.23              \\
      $a_{16}$  & $\qlz^{16}$             &    3.47\times10^{-5}      &    1.30\times10^{-3} & $d_6$    & $\qhz^6$        &   -7.24\times10^{-1} \\
      $a_{18}$  & $\qlz^{18}$             &                           &   -6.04\times10^{-5} & $d_8$    & $\qhz^8$        &    3.97\times10^{-1} \\
      $a_{20}$  & $\qlz^{20}$             &                           &    1.23\times10^{-6} & $d_{10}$ & $\qhz^{10}$     &   -1.38\times10^{-2} \\
      $c_4$     & $\qhx^4$                &   63.17                   &   78.60              & $d_{12}$ & $\qhz^{12}$     &    5.99\times10^{-5} \\
      $c_6$     & $\qhx^6$                &   -7.33\times10^{-1}      &   -1.00              & $v_1$    & $\qlz^3 \qhz$   &  119.42              \\
      $c_8$     & $\qhx^8$                &    4.38\times10^{-1}      &    7.22              & $v_2$    & $\qlz^2 \qhz^2$ &  212.45              \\
      $c_{10}$  & $\qhx^{10}$             &   -1.68\times10^{-2}      &   -3.79\times10^{-2} & $v_3$    & $\qlz \qhz^3$   &  169.37              \\
      $c_{12}$  & $\qhx^{12}$             &    1.29\times10^{-4}      &    6.49\times10^{-4} & $w_1$    & $\qlz^5 \qhz$   &    2.88              \\
      $l$       & $\qlz^2 \qhx^2$         &   -5.95                   &   -5.81              & $w_2$    & $\qlz^4 \qhz^2$ &   11.23              \\
      $m_1$     & $\qlz^4 \qhx^2$         &   -1.03                   &   -1.00              & $w_3$    & $\qlz^3 \qhz^3$ &   23.54              \\
      $m_2$     & $\qlz^2 \qhx^4$         &   -3.05                   &   -4.12              & $w_4$    & $\qlz^2 \qhz^4$ &   25.42              \\
      $n_1$     & $\qlz^4 \qhx^4$         &    1.85\times10^{-1}      &    2.41\times10^{-1} & $w_5$    & $\qlz \qhz^5$   &   13.46              \\
      $n_2$     & $\qlz^6 \qhx^2$         &    4.35\times10^{-3}      &    0.00              \\
      $n_3$     & $\qlz^2 \qhx^6$         &   -2.37\times10^{-1}      &   -3.14\times10^{-1} \\
      $t_1$     & $\qlx^3 \qhx$           & -118.35                   &   97.38              \\
      $t_2$     & $\qlx^2 \qhx^2$         &  215.00                   &  208.76              \\
      $t_3$     & $\qlx \qhx^3$           & -175.58                   &  195.22              \\
      $u_1$     & $\qlx^5 \qhx$           &   -2.72                   &    1.73              \\
      $u_2$     & $\qlx^4 \qhx^2$         &   10.64                   &    6.93              \\
      $u_3$     & $\qlx^3 \qhx^3$         &  -22.81                   &   18.34              \\
      $u_4$     & $\qlx^2 \qhx^4$         &   25.38                   &   24.27              \\
      $u_5$     & $\qlx \qhx^5$           &  -13.70                   &   15.57              \\
      $p$       & $\qlz^2 \qlx^2$         &    6.29                   &    6.02              \\
      $q_1$     & $\qlz^4 \qlx^2$         &   -1.70                   &   -1.56              \\
      $q_2$     & $\qlz^2 \qlx^4$         &   -1.70                   &   -1.39              \\
      $r_1$     & $\qlz^4 \qlx^4$         &    9.35\times10^{-2}      &    7.39\times10^{-2} \\
      $r_2$     & $\qlz^6 \qlx^2$         &    5.23\times10^{-3}      &    7.12\times10^{-3} \\
      $r_3$     & $\qlz^2 \qlx^6$         &    5.23\times10^{-3}      &    1.45\times10^{-2} \\
      $d$       & $\qlz^2 \qlx \qhx$      &  -19.09                   &   15.02              \\
      $e_1$     & $\qlz^2 \qlx^3 \qhx$    &    6.61                   &   -5.5               \\
      $e_2$     & $\qlz^2 \qlx^2 \qhx^2$  &  -13.16                   &  -13.62              \\
      $e_3$     & $\qlz^2 \qlx \qhx^3$    &   11.32                   &  -13.03              \\
      $f_1$     & $\qlz^2 \qlx^5 \qhx$    &    2.99\times10^{-1}      &   -2.06\times10^{-1} \\
      $f_2$     & $\qlz^2 \qlx^4 \qhx^2$  &   -6.80\times10^{-1}      &   -4.44\times10^{-1} \\
      $f_3$     & $\qlz^2 \qlx^3 \qhx^3$  &    1.39                   &   -1.10              \\
      $f_4$     & $\qlz^2 \qlx^2 \qhx^4$  &   -1.45                   &   -1.34              \\
      $f_5$     & $\qlz^2 \qlx \qhx^5$    &    8.75\times10^{-1}      &   -9.52\times10^{-1} \\
      $g$       & $\qlz^4 \qlx \qhx$      &    1.02                   &   -4.06\times10^{-1} \\
      $h_1$     & $\qlz^4 \qlx^3 \qhx$    &   -4.62\times10^{-1}      &    3.31\times10^{-1} \\
      $h_2$     & $\qlz^4 \qlx^2 \qhx^2$  &    8.13\times10^{-1}      &    7.51\times10^{-1} \\
      $h_3$     & $\qlz^4 \qlx \qhx^3$    &   -7.45\times10^{-1}      &    7.51\times10^{-1} \\
    \end{tabular}
  \end{ruledtabular}
\end{table*}

The nonharmonic potential $V^{\textrm{nh}}$ defined in the main text
is $V$ without the harmonic $\frac{1}{2} \Omega Q^2$ terms. The values
of all the coefficients in Eqs.~\ref{eq:full} and \ref{eq:qlzqhz}
obtained from a fit to the calculated energy surfaces of cubic and
strained KTaO$_3$ are given in Table~\ref{tab:coup-app}. I note that
values lower than the magnitude of $10^{-5}$ are below the accuracy of
the density functional calculations. They are kept so that that the
highest order anharmonicity has a positive sign, which keeps the
numerical solution of the equation of motions stable.
 
\section{Mode effective charges}
The mode effective charge vector $Z^*_{m,\alpha} = \partial
F_{m,\alpha}/\partial E_{\alpha}$ relates the force $F_{m,\alpha}$
experienced by the normal mode coordinate $Q_{m}$ due to an electric
field $E_{\alpha}$ along the direction $\alpha$. It is related to the
Born effective charges $Z^*_{\kappa,\alpha\beta}$ of atoms $\kappa$ in
the unit cell of a material by\cite{gonz97}
\begin{equation*}
  Z^*_{m,\alpha} = \sum_{\kappa,\beta}
  Z^*_{\kappa,\alpha\beta} U_m(\kappa,\beta),
\end{equation*}
where $U_m(\kappa,\beta)$ is the $\mathbf{q} = 0$ eigendisplacement
vector normalized as
\begin{equation*}
  \sum_{\kappa,\beta} M_{\kappa} [ U_m(\kappa,\beta)]^*
  U_n(\kappa,\beta) = \delta_{mn}. 
\end{equation*}
Here $M_{\kappa}$ is the mass of the atom $\kappa$. The
eigendisplacement vector is related to the eigenvector
$w_m(\kappa,\beta)$ of the dynamical matrix by
\begin{equation*}
  U_m(\kappa,\beta) = \frac{w_m(\kappa,\beta)}{\sqrt{M_\kappa}}.
\end{equation*}
Note that this definition of the mode effective charge is slightly
different from the one used in Ref.~\onlinecite{gonz97}. Here,
$Z^*_{m,\alpha}$ is related to the change in the value of the normal
mode coordinate rather than the change in the atomic displacements due
to a motion along the normal mode coordinate. This gives a different
normalization factor for $Z^*_{m,\alpha}$, and this quantity is
expressed in the units of $e$ amu$^{-\frac{1}{2}}$. Its sign is
arbitrary because the eigenvector of the dynamical matrix is defined
up to a multiplicative constant.

The mode effective charge can be experimentally determined. It is
related to the ionic contribution to the dielectric constant by\cite{gonz97}
\begin{equation*}
  \epsilon_{\alpha\beta}(\Omega) = \epsilon^{\infty}_{\alpha\beta} +
  \frac{4\pi}{V_0} \sum_m \frac{Z^*_{m,\alpha}
    Z^*_{m,\beta}}{\Omega^2_m - \Omega^2},
\end{equation*}
where $V_0$ is the unit cell volume and $\Omega_m$ is the frequency of
the mode $m$. This expression shows that the oscillator strength
measured in optical spectroscopy is the square of the mode effective
charge.



\begin{thebibliography}{99}
\bibitem{q1q2q3} {Ref.~\onlinecite{jura16} claims that a $\qr
  Q_{\textrm{IR}_1}Q_{\textrm{IR}_2}$ (with 1 $\neq$ 2) coupling
  causes a rectification along $\qr$. However, such a coupling
  actually leads to an oscillation of the $\qr$ mode with an envelop
  frequency $\Omega_{\textrm{IR}_1} - \Omega_{\textrm{IR}_2}$.}
\end{thebibliography}
\end{document}